# PRECISE COMPUTATION OF POSITION ACCURACY IN GNSS SYSTEMS

by Juan Pablo Boyero Garrido
(jpboyero@gmv.es)

**ABSTRACT**

Accuracy and Availability computations for a GNSS System – or combination of Systems – through Service Volume Simulations take considerable time. Therefore, the computation of the accuracy in 2D and 3D are often simplified by an approximate solution. The drawback is that such simplifications can lead to accuracy results that are too conservative (up to 25% in the 2D case and up to 43% in the 3D case, for a 95% confidence level), which in turn translates into pessimistic System Availability.

This article presents a way to compute the exact accuracy, for any confidence level, for the one, two or three dimensional cases, through the derivation of corresponding factors. Using the factors introduced here, allows getting accurate results swiftly.

The generic mathematical solution to compute the accuracy is presented. The approximate and precise computations are described. Then, the exact factors that should be applied to obtain the accuracy at a typical confidence level (95%) are derived for the three cases.

## 1. INTRODUCTION

In a Satellite Navigation System, the position errors in the user domain can be computed for the one, two or three dimensional cases, and the Availability of the system is then evaluated as the percentage of time during which the accuracy is better than the specified value. The same is valid for the velocity errors.

In Service Volume Simulations, position errors in the user domain at each epoch are usually modeled in every component as Gaussian random variables. This derives, in turn, from the modeling of the satellites ranging accuracy errors as independent Gaussian random variables.

Starting from the components of the position domain error at a given location and instant, the one, two or three dimensional errors can be computed at the specified confidence level.

For instance, sometimes the accuracy in 3D is given in terms of SEP (Spherical Error Probable), which represents the radius of the sphere containing the error with 50% probability. Other times it is more convenient to work with different confidence levels – 95% is another typical value. (see [1] for a list of commonly used measures and their definitions).

In practice, the simulations run to compute the accuracy and Availability for a GNSS System – or combination of Systems - take considerable time. That is why the computation of the accuracy in 2D and 3D are often simplified to reduce the computational load.

The drawback is that such simplifications can lead to accuracy results that are too conservative and consequently imply pessimistic results in terms of System Availability.

In turn, working with pessimistic Availability results can have a relevant impact when designing a System, since improvement of Availability requires either to trade off other System performance parameters, or to re-dimension some System components (such as the ground segment).

This article presents a way to compute the exact accuracy for any confidence level for the one, two or three dimensional cases, through the derivation of corresponding factors. Using the factors introduced here - as look up table for instance – allows getting accurate results swiftly.

First, the generic mathematical solution to compute the accuracy is presented. Then, the exact factors that should be applied to obtain the accuracy at a given confidence level (95%) are derived for the cases: monodimensional (single value), bidimensional (curve) and three dimensional (surface).

It is fair to recall that the generic mathematical solution is also presented in [2]. For instance, the 2D case is treated with great detail in Annex D of that reference. The perspective offered here is, however, different.

In [2] the development of the generic mathematical solution is oriented in such a way that the final integral to obtain a specific factor can be solved through interpolation in corresponding tables. Taking advantage of the computational power now available, the development made below is oriented towards expressions which are later on quickly solved through numerical integration.

The idea behind is to allow for the fast computation of the whole range of exact factors needed, factors which can then be directly applied to GNSS Systems simulations, and which can also help in quantifying the inaccuracies associated to the simplifications in usual simulations.

This is more evident for the 3D case, where additional development has been done here, leading to the computation of the whole range of significant factors (surface), from which relevant conclusions have also been drawn.

## 2. MONODIMENSIONAL CASE

The monodimensional is the usual case for the computation of the Vertical accuracy. The position domain error in the vertical dimension is described by a Gaussian random variable with zero mean and standard deviation $\sigma$. The standard deviation comes directly from the third element in the diagonal of the navigation solution covariance matrix.

Hence, the probability density function of the vertical error ($x$) is:

$$f(x) = \frac{1}{\sigma\sqrt{2\pi}} \exp(\frac{-x^2}{2\sigma^2})$$

**Eq. 1**

As shown in the next figure, the accuracy related to a certain confidence level is the value ($e$) for which the probability $p(e)$ equals that level.

$$p(e) = P(-e < x < e) = \int_{-e}^{e} f(x) \cdot dx$$

**Eq. 2**

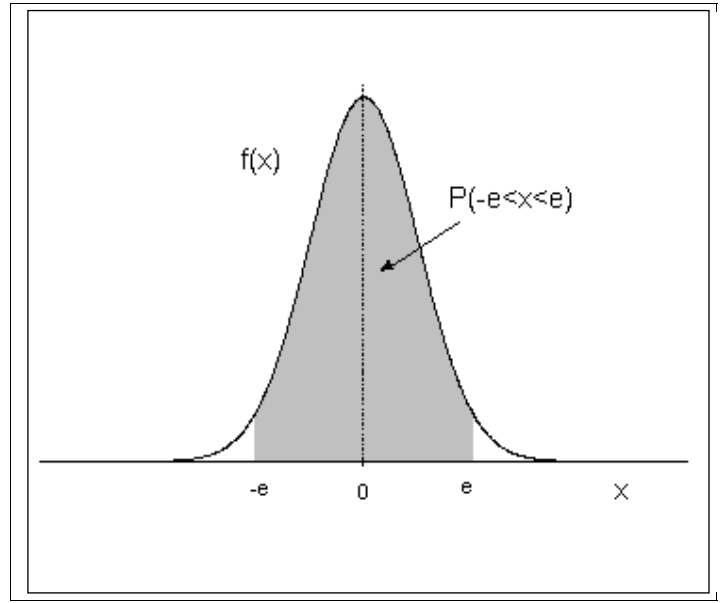

**Figure 1: Probability density function and accuracy for a given confidence level in the monodimensional case**

The error function can then be used to compute the exact factor for the desired confidence level:

$$p(e) = \int_{-e}^{e} f(x) \cdot dx = erf(\frac{e}{\sigma\sqrt{2}}) \qquad \text{Eq. 3}$$

For example, for a Vertical accuracy with 95% confidence level for, p=0.95 $\Rightarrow$ e=1.96$\sigma$, the factor should be 1.96 applied.

## 3. BIDIMENSIONAL CASE

The bidimensional is the usual case for the computation of the Horizontal accuracy. The initial point is now the covariance matrix of the position domain errors in the horizontal plane, that is, the upper-left 2x2 submatrix of the navigation solution covariance matrix.

### 1.1 APPROXIMATE COMPUTATION IN THE BIDIMENSIONAL CASE

Two main lines are usually followed for the computation in the 2D case.

The simplest one consists of considering only the elements in the diagonal of the matrix. The Horizontal accuracy can be directly computed as the square root the sum of the two elements in the diagonal. This is the same operation performed in the computation of the HDOP, that is, assuming UERE equal to 1 meter for all satellites, the Horizontal accuracy would be equal to HDOP.

A factor can then be applied to convert such value into the value corresponding to the desired level of confidence. Specifically, for a 95% confidence level the factor habitually applied is 1.96.

The second line is in turn based on the eigenvalues of the 2x2 submatrix. By computing the eigenvalues, the variances of the errors in orthogonal directions are found. Moreover, the largest eigenvalue corresponds to the variance of the error along the worst direction in the plane. The situation is depicted in Figure 2, where (x´,y´) represent the dimensions to which the covariance matrix is referred, while the dimensions (x,y) represent those relative to the eigenvalues.

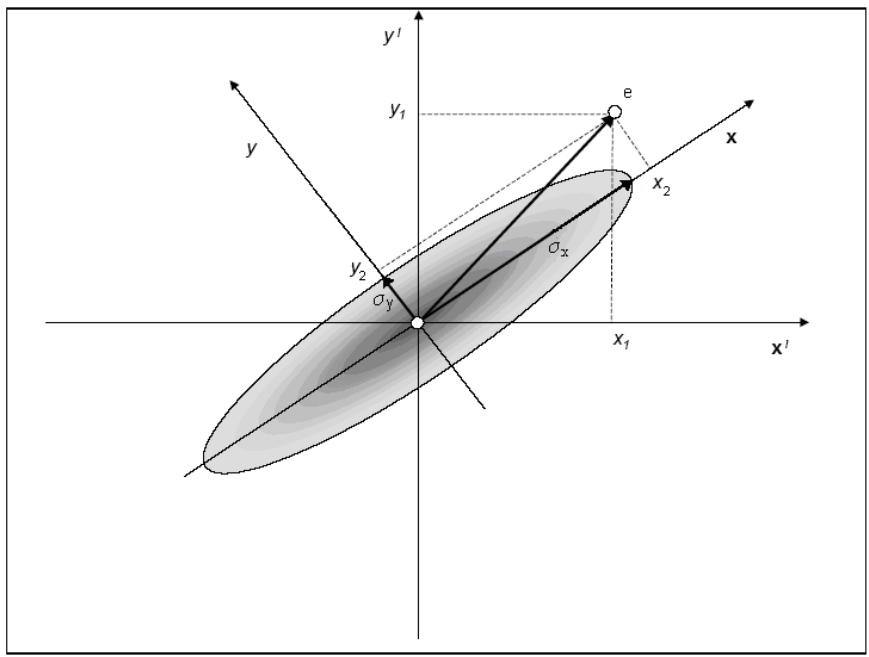

**Figure 2: Error ellipse in the horizontal plane**

The squared error in the horizontal plane can then be obtained as

$$e^2 = x_2^2 + y_2^2$$ **Eq. 4**

Where *x* and *y* are Gaussian random variables with zero mean and standard deviations $\sigma_x$ and $\sigma_y$ respectively. They determine the error ellipse, as shown in Figure 2.

At this stage, in order to simplify the computation of the accuracy at a given level of confidence, an approximation is normally done. The idea is to assume that both variances of the error in the two orthogonal directions are equal to the largest one (e.g. $\sigma=\sigma_x$). With this conservative simplification, the squared error becomes a Chi-Squared random variable with 2 degrees of freedom (see [3], p187).

The Chi-Squared distribution is available in most mathematical software tools. Alternatively, it can also be found in tabulated form. Therefore, with the described approximation it is quite easy to obtain the factor to convert the value $\sigma_x$ into that corresponding to the desired level of confidence. Specifically, for a 95% confidence level the factor is 2.447.

Thus, the usual method to compute the Horizontal accuracy based on eigenvalues consists of 1) computing the eigenvalues of the 2x2 submatrix of the navigation solution covariance matrix; 2) retaining the maximum eigenvalue, which corresponds to the variance of the error along the worst direction in the plane (σ); and 3) multiplying the standard deviation by the factor corresponding to the desired confidence level (e.g. 2.447σ for 95%).
The computation is depicted in Figure 3 below.

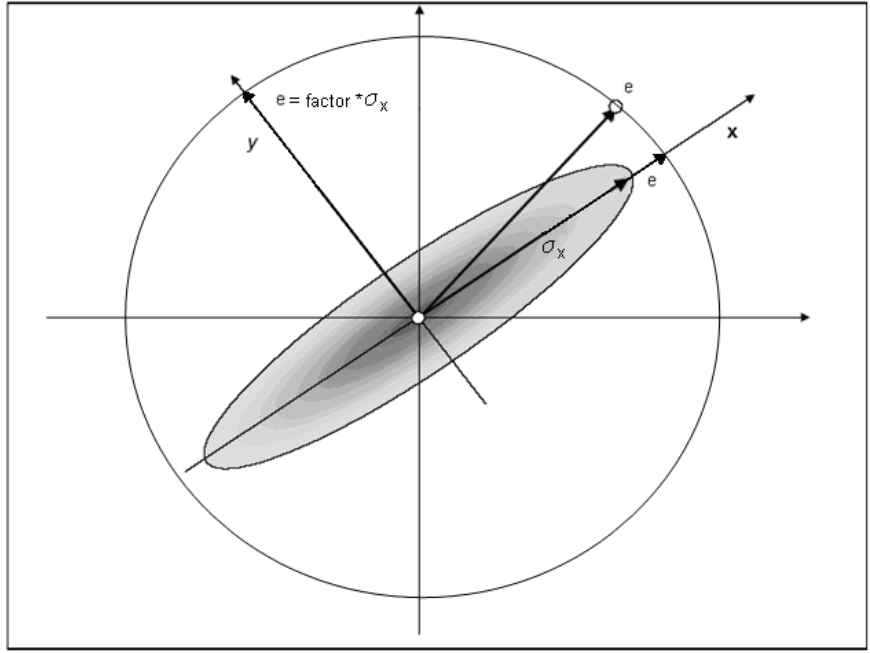

**Figure 3: Factor to obtain the error at a given confidence level from the largest variance in a two dimensional Gaussian distribution**

## 1.2 PRECISE COMPUTATION IN THE BIDIMENSIONAL CASE

In order to find the exact value linked to a given confidence level, we will start from the eigenvalues of the 2x2 submatrix of the navigation solution covariance matrix. Since the eigenvalues are the variances of two independent Gaussian distributions with zero mean, we can form the joint distribution $j(x,y)$ as:

$$j(x,y) = f_x(x) \cdot f_y(y) = \frac{1}{\sigma_x \sqrt{2\pi}} \exp(\frac{-x^2}{2\sigma_x^2}) \cdot \frac{1}{\sigma_y \sqrt{2\pi}} \exp(\frac{-y^2}{2\sigma_y^2})$$

Eq. 5

The distribution is sketched in Figure 4. The lower part of the figure shows increasing ellipse errors, in line with the one shown in Figure 2 above. From this perspective, the approximation of assuming that both variances of the error in the two orthogonal directions are equal to the largest, can be seen as a substitution of the actual joint distribution by another one having a more spread probability density function.

Through corresponding double integral, the two dimensional joint distribution allows for the derivation of the probability of the errors associated to a specific region, $p$:

$$p = \int_y \int_x j(x,y) \cdot dx \cdot dy$$

Eq. 6

Providing that we are interested in the error (e) that is not surpassed with the given level of confidence, the integral in Eq. 6 must sweep the region given by a circle of radius *e*. In polar coordinates, this region is defined by $\varphi \in [0, 2\pi)$ and $r \in [0, e)$, and the integral is expressed as (applying $x = r\cos\varphi$ and $y = r\sin\varphi$):

$$p(e) = \int_0^{2\pi} \int_0^e j(r\cos\varphi, r\sin\varphi) \cdot dr \cdot r d\varphi \qquad \text{Eq. 7}$$

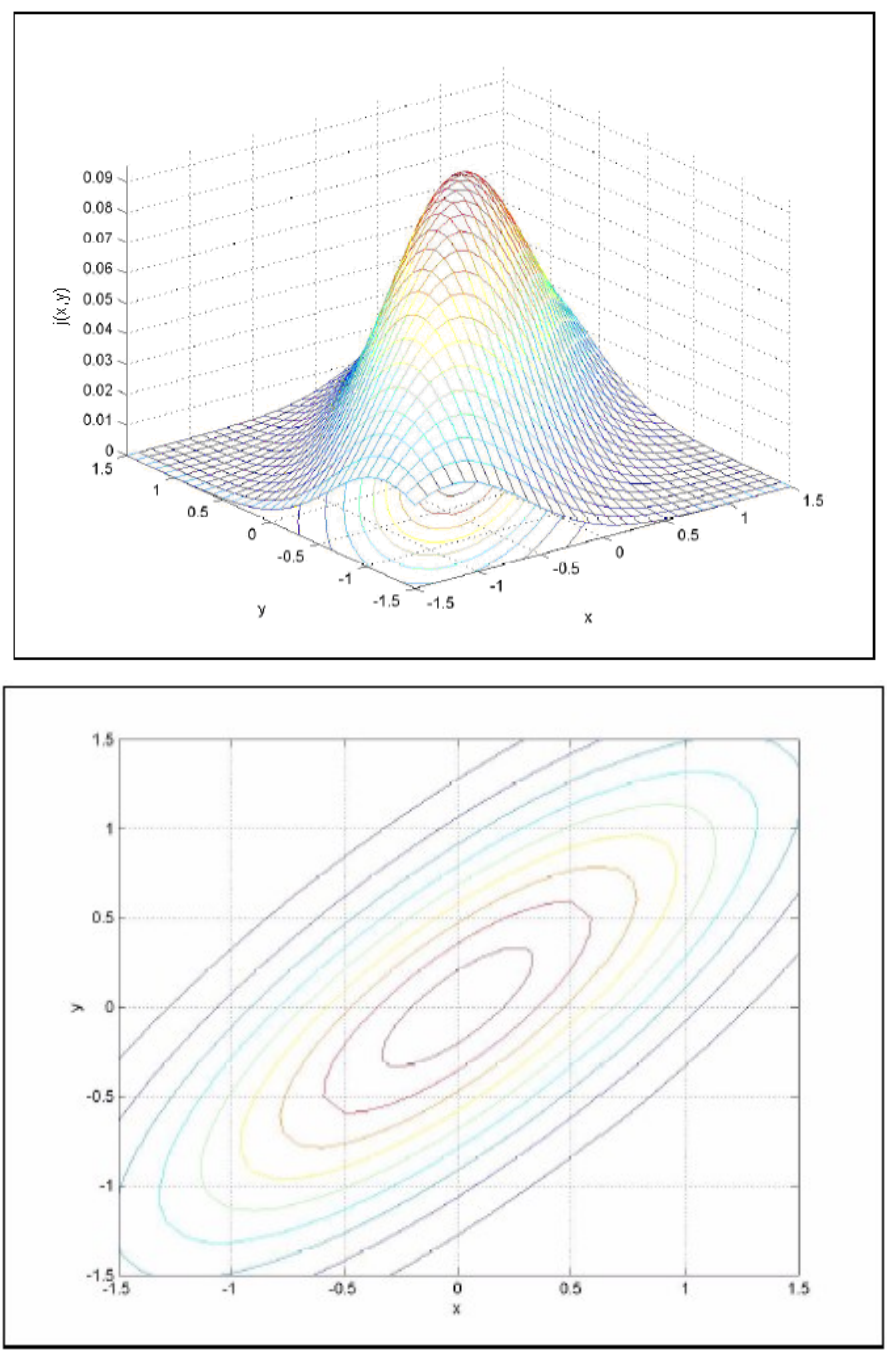

**Figure 4: Joint distribution of two independent Gaussian random variables with zero mean**

In the particular cases where the standard deviation is the same for both distributions - mathematically $\sigma_x = \sigma_y = \sigma$ - the integral becomes

$$p(e) = \int_0^{2\pi} \int_0^e \frac{1}{2\pi\sigma^2} \exp\left(\frac{-r^2}{2\sigma^2}\right) \cdot dr \cdot r d\varphi =$$

$$= \frac{1}{\sigma^2} \int_0^e \exp\left(\frac{-r^2}{2\sigma^2}\right) \cdot r dr =$$

$$= -\exp\left(\frac{-r^2}{2\sigma^2}\right)\Bigg|_0^e = 1 - \exp\left(\frac{-e^2}{2\sigma^2}\right)$$

Eq. 8

This again corresponds to the Chi-Squared distribution with 2 degrees of freedom, from which *e* can be analytically derived as:

$$\boxed{e = \sqrt{-2\sigma^2 \ln(1-p)} = \sigma\sqrt{-2\ln(1-p)}}$$

Eq. 9

For 95% confidence, $p=0.95 \Rightarrow e=2.447\sigma$, the same result presented above.

In a general case, however, the integral to be solved is

$$p(e) = \int_0^{2\pi} \int_0^e \frac{1}{2\pi\sigma_x \sigma_y} \exp\left[\frac{-r^2}{2}\left(\frac{\cos^2\varphi}{\sigma_x^2} + \frac{\sin^2\varphi}{\sigma_y^2}\right)\right] \cdot dr \cdot r d\varphi =$$

Eq. 10

$$= \frac{1}{2\pi\sigma_x\sigma_y} \int_0^{2\pi} \left(\int_0^e \exp\left[\frac{-r^2}{2}\left(\frac{\cos^2\varphi}{\sigma_x^2} + \frac{\sin^2\varphi}{\sigma_y^2}\right)\right] \cdot r \cdot dr\right) \cdot d\varphi =$$

$$= \frac{1}{2\pi\sigma_x\sigma_y} \int_0^{2\pi} \left(\frac{-1}{\left[\frac{\cos^2\varphi}{\sigma_x^2} + \frac{\sin^2\varphi}{\sigma_y^2}\right]} \exp\left[\frac{-r^2}{2}\left(\frac{\cos^2\varphi}{\sigma_x^2} + \frac{\sin^2\varphi}{\sigma_y^2}\right)\right]\right)\Bigg|_0^e \cdot d\varphi =$$

$$= \frac{1}{2\pi\sigma_x\sigma_y} \int_0^{2\pi} \frac{1}{\left[\frac{\cos^2\varphi}{\sigma_x^2} + \frac{\sin^2\varphi}{\sigma_y^2}\right]} \cdot d\varphi - \frac{1}{2\pi\sigma_x\sigma_y} \int_0^{2\pi} \frac{1}{\left[\frac{\cos^2\varphi}{\sigma_x^2} + \frac{\sin^2\varphi}{\sigma_y^2}\right]} \exp\left[\frac{-e^2}{2}\left(\frac{\cos^2\varphi}{\sigma_x^2} + \frac{\sin^2\varphi}{\sigma_y^2}\right)\right] \cdot d\varphi$$

The first term in the previous expression becomes equal to 1. One way to see this is by taking advantage of the fact that *p(e)* is a joint distribution function, and therefore, must be normalized to 1. Mathematically this means that is $p \rightarrow 1$ when $e \rightarrow \infty$. Since, on the other hand, *p(e)* tends precisely to the expression in the first term:

$$p(e) \rightarrow \frac{1}{2\pi\sigma_x\sigma_y} \int_0^{2\pi} \frac{1}{\left[\frac{\cos^2\varphi}{\sigma_x^2} + \frac{\sin^2\varphi}{\sigma_y^2}\right]} \cdot d\varphi$$

Eq. 11

We can thus convert Eq. 10 into:

$$p(e) = 1 - \frac{1}{2\pi\sigma_x\sigma_y} \int_0^{2\pi} \frac{1}{\left[\frac{\cos^2\varphi}{\sigma_x^2} + \frac{\sin^2\varphi}{\sigma_y^2}\right]} \exp\left[\frac{-e^2}{2}\left(\frac{\cos^2\varphi}{\sigma_x^2} + \frac{\sin^2\varphi}{\sigma_y^2}\right)\right] \cdot d\varphi \qquad \textbf{Eq. 12}$$

Let´s now introduce a parameter $v = \sigma_x^2/\sigma_y^2$, which is the ratio between the largest and smallest variances. This parameter allows writing the equation as a function of just $e/\sigma_x$ and $v$ itself:

$$\boxed{1 - p(e) = \frac{\sqrt{v}}{2\pi} \int_0^{2\pi} \frac{1}{\left[\cos^2\varphi + v\sin^2\varphi\right]} \exp\left[\frac{-e^2}{2\sigma_x^2}(\cos^2\varphi + v\sin^2\varphi)\right] \cdot d\varphi} \qquad \textbf{Eq. 13}$$

Recall that both $\sigma_x$ (the largest standard deviation) and $v$ are known as soon as the eigenvalues of the covariance matrix are available, that is, from the beginning of this problem.

For a given value of $p$, the resolution of Eq. 13 allows obtaining the factor $e/\sigma_x$ as a function of $v$. For instance, for $v=1$ (which is the case when $\sigma_x=\sigma_y$) and p=95, the factor obtained through this formula is, as expected, 2.447.

For the results presented here, the formula has been solved numerically by means of Matlab (own code; see also [4] for online available code to compute $p$ taking as input the covariance matrix).

Conversely, the factor by which to multiply the largest standard deviation ($\sigma_x$) can also be derived as a function of the inverse of the square root parameter $v$ previously defined $r=1/\sqrt{v}$ (as well with $1/v$). That is, instead of dividing the largest variance by the smallest one, dividing the smallest standard deviation by the largest one.

Although this way may create some problems with the computations when the value of the abscissa tends to zero (computations which can be made separately with the parameter $v$ if needed), it has the advantage of presenting a more compact plot, since the range $r=1/\sqrt{v}$ goes only from 0 to 1.

Following that line, Figure 5 displays the factor for the bidimensional case as a function of the ratio smallest/largest standard deviation $r$, specifically for the typical 95% confidence level (up), as well as for other percentages (down).

The curve for the 50% confidence level is included in Figure 5, and it can be checked that it matches the curve in Figure 8 from [2], where the same values are computed using a different method (general integral is developed in a different manner so that it can be solved by interpolation of relevant tables).

From Figure 5 it can be seen as well that the exact factor follows for all percentages a function which increases monotonously with the ratio smallest/largest standard deviation.

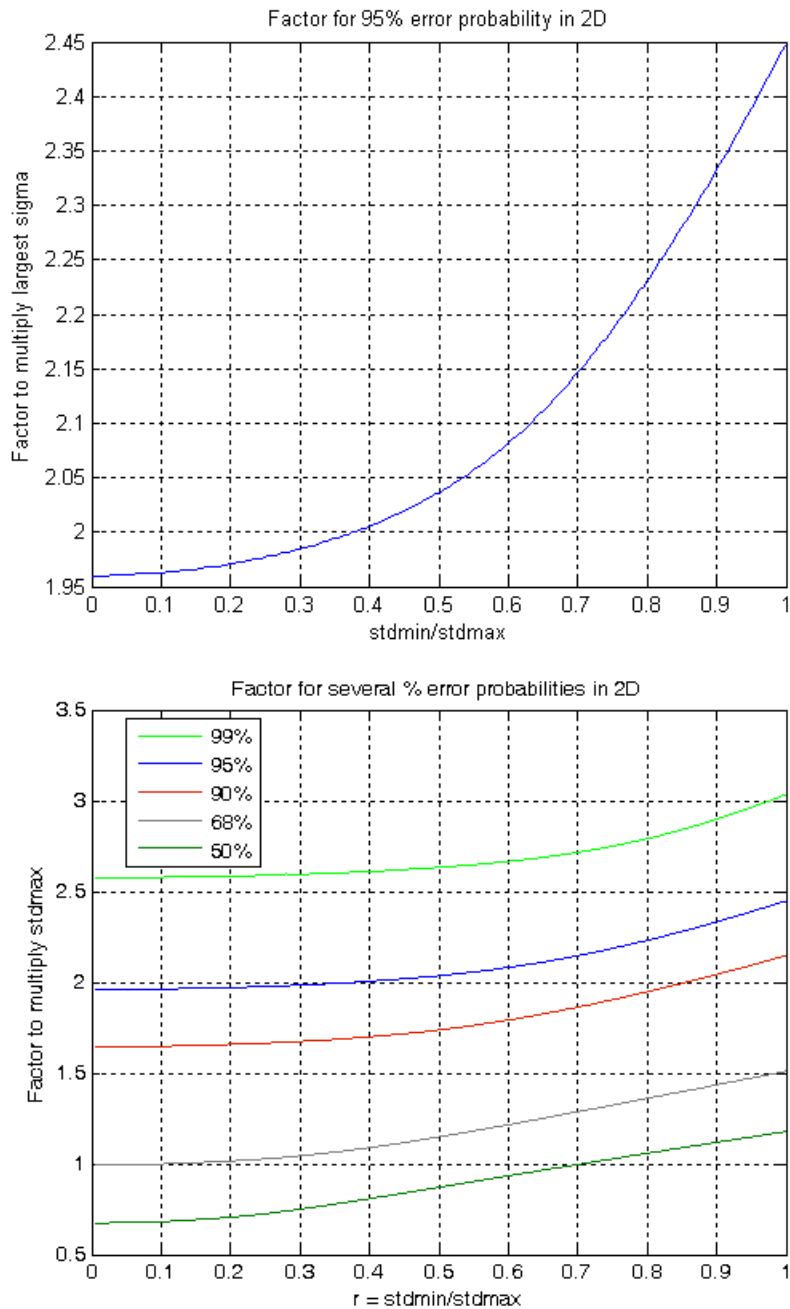

**Figure 5: Factor to multiply the largest std to obtain the 2D error, as a function of the ratio stdmin/stdmax, for typical 95% confidence (up) and for several % (down)**

The exact factor for the typical 95% confidence level is compared in the next figure against the factor which multiplies the largest standard deviation in the two simplified approaches described above, namely, the approximate approach with eigenvalues but using a unique single factor (2.45); and the approach based on the square root the sum of the two elements in the diagonal of the 2x2 submatrix multiplied by 1.96.

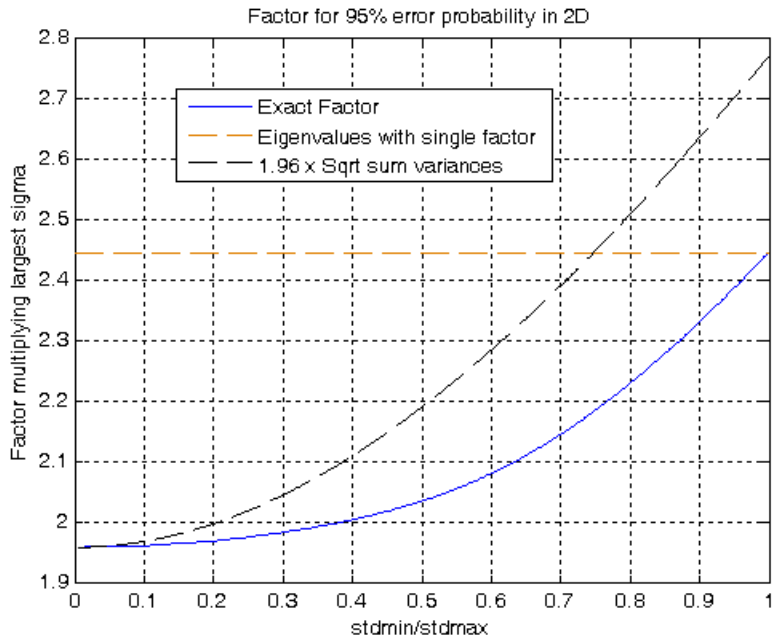

**Figure 6: Comparison of factor multiplying the largest std to obtain the 2D error at 95% confidence as a function of the ration stdmin/stdmax**

It becomes visible in Figure 6 that the approximate approach of using a unique single factor (2.45) to compute the Horizontal accuracy can lead to a considerable overestimation of the error.
For instance, taking an intermediate point, $v$=3 ($r$ =0.58) that approach represents an error increment of 18.2% with respect to the exact value (2.45/2.07 = 1.182).

The maximum of the overestimation occurs for the case when $v \rightarrow \infty$ ($r \rightarrow 0$), that is, when the error in one of the dimensions is clearly dominant, and the factor for p=0.95 tends to 1.96, the one obtained in the monodimensional case. In that case, the overestimation is equal to (2.45-1.96)/1.96 = 25%.

In the same manner, the approach based on the square root the sum of the two elements in the diagonal of the 2x2 submatrix also leads to overestimation.
With such approach, the maximum of the overestimation occurs for the case when $r \rightarrow 1$ – there is no dominant dimension – since $1.96 \cdot \sqrt{(2\sigma^2_x)} = 2.77\sigma_x$, with overestimation of (2.77-2.45)/2.45 = 13%.

## 4. THREE DIMENSIONAL CASE

For the three dimensional case it is quite simple to establish a parallelism with the two dimensional case, the starting point being the covariance matrix of the position domain errors in the three dimensional space, that is, the upper-left 3x3 submatrix of the position solution covariance matrix.
The same reasoning is valid for the velocity errors, provided that the covariance matrix for the velocity solution is built in the same way as the one for the position solution, the only difference being the use of range rates instead of UERE´s.

## 1.3 APPROXIMATE COMPUTATION IN THE THREE DIMENSIONAL CASE

The same two options showed in the two dimensional case are followed in this case. The one considering just the elements in the diagonal of the matrix computes the accuracy as the square root the sum of the three elements (similar to the PDOP computation).

The second option is based on the eigenvalues of the 3x3 submatrix. By computing the eigenvalues, the variances of the errors in three orthogonal directions are found. The squared error in the horizontal plane can be then obtained as

$$e^2 = x_2^2 + y_2^2 + z_2^2$$  **Eq. 14**

Where *x*, *y* and *z* are Gaussian random variables with zero mean and standard deviations $\sigma_x$, $\sigma_y$ and $\sigma_z$ respectively. They determine the error ellipsoid, where the largest eigenvalue corresponds to the variance of the error along the worst direction in the three dimensional space.

After doing the same simplification of assuming that the variances of the errors in the three orthogonal directions are equal to the largest one, the squared error becomes a Chi-Squared random variable with 3 degrees of freedom.

From the Chi-Squared distribution, the factor corresponding to the desired confidence level can be obtained. Hence, for the 95% confidence level, the factor for the 3D space becomes 2.795. Consequently, the formula applied for the error is e=2.795$\sigma$ ($\sigma$ being the largest eigenvalue from the 3x3 covariance matrix).

## 1.4 PRECISE COMPUTATION IN THE THREE DIMENSIONAL CASE

In order to derive the exact accuracy value for a given confidence level in the three dimensional case, the derivation yielding to Eq. 5 needs to be extended. With the three eigenvalues, representing the variances of three independent Gaussian distributions with zero mean, we can form the joint distribution *j(x,y,z)* as:

$$j(x,y,z) = f_x(x) \cdot f_y(y) \cdot f_z(z) = \frac{1}{\sigma_x \sqrt{2\pi}} \exp(\frac{-x^2}{2\sigma_x^2}) \cdot \frac{1}{\sigma_y \sqrt{2\pi}} \exp(\frac{-y^2}{2\sigma_y^2}) \cdot \frac{1}{\sigma_z \sqrt{2\pi}} \exp(\frac{-z^2}{2\sigma_z^2})$$  **Eq. 15**

Now it is a triple integral the one which solves the problem. The probability *p* of the errors associated to a specific region is determined by:

$$p = \int_z \int_y \int_x j(x,y,z) \cdot dx \cdot dy \cdot dz$$  **Eq. 16**

To illustrate the analogy with the bidimensional case, we can work out the counterpart of Eq. 8, namely, the particular case where the standard deviation is the same for the three distributions, $\sigma_x=\sigma_y=\sigma_z=\sigma$. The joint distribution turns into:

$$j(x,y,z) = \left(\frac{1}{\sigma\sqrt{2\pi}}\right)^3 \exp(-\frac{x^2+y^2+z^2}{2\sigma^2})$$  **Eq. 17**

In polar coordinates (using $x = r\sin\theta\cos\varphi$, $y = r\sin\theta\sin\varphi$ and $z = r\cos\theta$), the integral has to be done over the polar variables $\theta \in [0,\pi]$, $\varphi \in [0,2\pi)$ and $r \in [0,e]$:

$$p(e) = \int_0^\pi \int_0^{2\pi} \int_0^e \left(\frac{1}{\sigma\sqrt{2\pi}}\right)^3 \exp\left(\frac{-r^2}{2\sigma^2}\right) \cdot dr \cdot rd\varphi \cdot r\sin\theta \cdot d\theta = \quad \text{Eq. 18}$$

$$= \frac{2}{\left(\sigma\sqrt{2\pi}\right)^3} \int_0^{2\pi} \int_0^e \exp\left(\frac{-r^2}{2\sigma^2}\right) \cdot dr \cdot r^2 d\varphi = \frac{4\pi}{\left(\sigma\sqrt{2\pi}\right)^3} \int_0^e \exp\left(\frac{-r^2}{2\sigma^2}\right) \cdot r^2 \cdot dr =$$

$$= \frac{4}{\sigma\sqrt{2\pi}} \int_0^e \exp\left(\frac{-r^2}{2\sigma^2}\right) \cdot \frac{r^2}{2\sigma^2} \cdot dr$$

Applying the change of variable $t = r^2/2\sigma^2$, which implies $dt = rdr/\sigma^2$, and changing the integration limits accordingly,

$$\boxed{p(e) = \frac{2}{\sqrt{\pi}} \int_0^{\frac{e^2}{2\sigma^2}} \exp(-t) \cdot t^{1/2} \cdot dt} \quad \text{Eq. 19}$$

This integral falls into the category of the so called incomplete gamma function[1]

$$\gamma(a,x) = \int_0^x \exp(-t) \cdot t^{a-1} \cdot dt \quad \text{Eq. 20}$$

In this case, $a = 3/2$; $x = e^2/2\sigma^2$. Therefore, the problem can be solved by finding the value for which

$$\gamma(3/2, x) = \frac{\sqrt{\pi}}{2} p \quad \text{Eq. 21}$$

The incomplete gamma function is as well implemented in most mathematical software tools (such as Matlab), or can otherwise be found tabulated. Through these means, it can be derived that for p=0.95, the corresponding $x \cong 3.908$, which in turn leads to $e \cong 2.795\sigma$, in line with the above result.

In the general three dimensional case, the triple integral in polar coordinates has the form:

$$p(e) = \frac{1}{\sigma_x \cdot \sigma_y \cdot \sigma_z} \left(\frac{1}{\sqrt{2\pi}}\right)^3 \int_0^\pi \int_0^{2\pi} \int_0^e \exp\left(\frac{-r^2}{2}\left[\left(\frac{\sin\theta\cos\varphi}{\sigma_x}\right)^2 + \left(\frac{\sin\theta\sin\varphi}{\sigma_y}\right)^2 + \left(\frac{\cos\theta}{\sigma_z}\right)^2\right]\right) \cdot dr \cdot rd\varphi \cdot r\sin\theta \cdot d\theta \quad \text{Eq. 22}$$

Taking the direction of x as the one corresponding to the largest eigenvalue, two parameters $m = \sigma_y/\sigma_x$ and $n = \sigma_z/\sigma_x$ can be defined, both ranging from 0 to 1. Notice that the direction of x is determined by the largest eigenvalue, and the directions y and z follow a trirectangular trihedron accordingly.

---

[1] In some references (see [5], p260), the term incomplete gamma function is applied to the normalized function $P(a,x) = \frac{\gamma(a,x)}{\Gamma(a)} = \frac{1}{\Gamma(a)} \int_0^x \exp(-t) \cdot t^{a-1} \cdot dt$

Applying then the change of variable $t = r/\sigma_x$, which implies $dt = dr/\sigma_x$, and changing the integration limits accordingly, the integral becomes:

$$p(e) = \frac{1}{m \cdot n}\left(\frac{1}{\sqrt{2\pi}}\right)^3 \int_0^\pi \int_0^{2\pi} \int_0^{e/\sigma_x} \exp\left(\frac{-t^2}{2}\left[(\sin\theta\cos\varphi)^2 + \left(\frac{\sin\theta\sin\varphi}{m}\right)^2 + \left(\frac{\cos\theta}{n}\right)^2\right]\right) \cdot \sin\theta \cdot t^2 \cdot dt \cdot d\varphi \cdot d\theta \qquad \textbf{Eq. 23}$$

It can be seen that the integral depends on the two parameters *m*, *n*, as well as on the factor $e/\sigma_x$.

The factor for p=0.95 has been numerically found here for each combination of *m*, *n*, the outcome being the figure below.

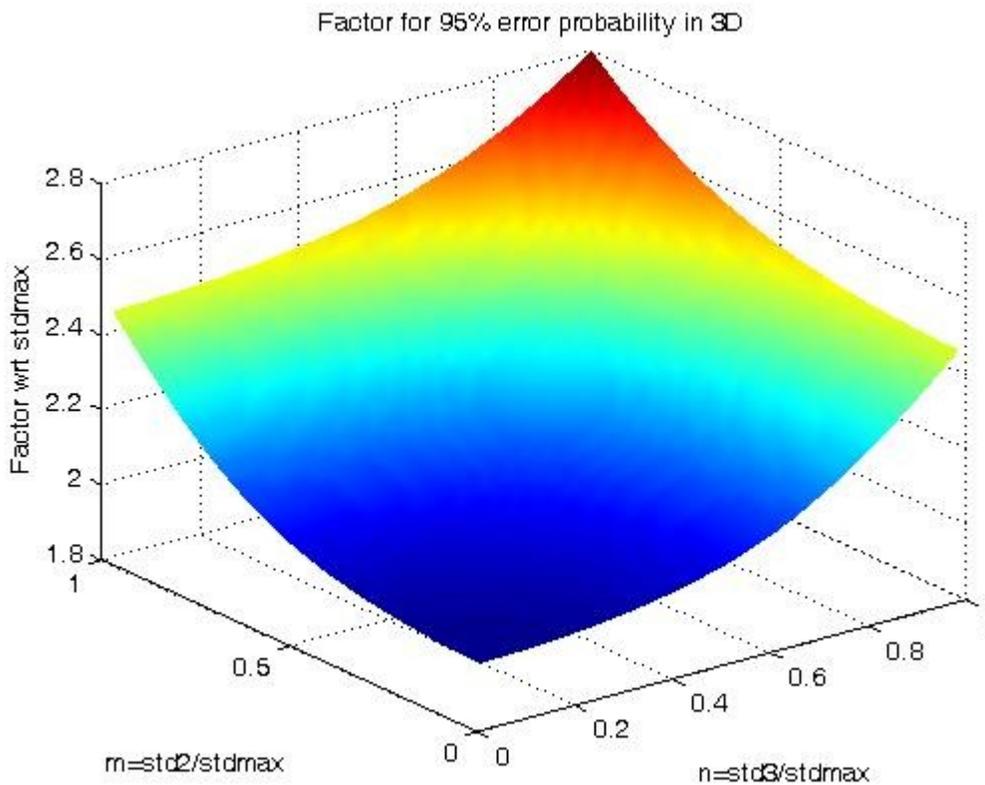

**Figure 7: Factor to multiply the largest std to obtain the error at 95% confidence in a three dimensional Gaussian distribution**

Figure 7 shows the factor by which the largest std has to be multiplied in order to find the radius corresponding to the 95% of the error in 3D.

The factor is presented as a function of the ratios between each of the other standard deviations and the largest standard deviation. From the Figure, it can be observed that the integral is symmetrical in *m*, *n*. This implies that, once the square root of the largest eigenvalue is taken to get $\sigma_x$, the other two eigenvalues can be interchangeably selected to obtain *m* and *n*.

As a cross check, when the parameters *m* and *n* are equal to 0, which means there is only one relevant dimension, the factor becomes 1.96, in line with the monodimensional case.

Conversely, when the three standard deviations are equal (*m*=*n*=1), the factor becomes $\cong$2.795; in line with the above result.

And when one of the standard deviations tends to 0, the surface reduces to a curve which matches the one in Figure 5.

In terms of overestimation, Figure 7 makes comprehensible the difference when using only the factor derived in the approximate approach.

For instance, the overestimation is in the order of 14% for cases where two of the errors are similar and dominant compared to the third one, that is *m*=1, *n*=0. (2.8/2.45=1.14).

The overestimation of the accuracy error in 3D can be as high as 43%, corresponding to the cases where the error in one direction is dominant with respect to the other two (*m*=*n*=0; 2.8/1.96=1.43).

In the same way, the approach based on the square root the sum of the two elements in the diagonal of the 2x2 submatrix also leads to overestimation.

With such approach for the 95% confidence level computation in 3D, the maximum of the overestimation occurs for the case when *m*=*n*=1 – there is no dominant dimension – since $1.96 \cdot \sqrt{(3\sigma^2_x)}$ = $3.395\sigma_x$, with overestimation of (3.395-2.795)/2.795 = 21.5%.

## 5. SUMMARY

This article presents a way to compute the exact position accuracy for GNSS Systems, for any desired confidence level. The results are valid as well for velocity accuracy.

In practice, the accuracy and Availability computations for a GNSS System are done through Service Volume Simulations, which take a long time, so that the computation of the accuracy in 2D and 3D are often simplified to reduce the computational load.

It has been shown here that such simplifications can lead to accuracy results that are too conservative (up to a 25% in the 2D case and up to a 43% in the 3D case), and consequently imply pessimistic results in terms of System Availability

The generic mathematical solution to compute the accuracy for the one, two and three dimensional cases has been derived - the first two cases being normally linked to the Vertical and Horizontal accuracies respectively.

From the generic mathematical solution, the exact factors that can be applied to obtain the accuracy at any desired confidence level may be computed.

The factors have been computed for the three cases: monodimensional (single value), bidimensional (curve) and three dimensional (surface), for a typical confidence level (95%).

By using these factors, the accuracy – and consequently the Availability – of a GNSS System can be obtained in a fast and correct manner.

Although the details of the implementation of the factors are not explicitly discussed in this article, taking for example the 2D case, where the factors follow a curve, easy implementations can be devised in the form of a look-up table, or in the form of a function fitting the curve which can in turn be quickly evaluated.

**REFERENCES**

1. [Frank van Diggelen](#) "GPS Accuracy: Lies, Damn Lies, and Statistics", GPS World, November 29, 1998
2. Aeronautical Chart and Information Center, USAF, Technical Report No. 96 "Principles of Error Theory and Cartographic Applications", February 1962
3. Athanasios Papoulis, Probability, Random Variables and Stochastic Processes, McGraw-Hill, Second Edition.
4. Davis, T. and Kleder, M. "Confidence region radius.", Mathworks Central File Exchange. (http://www.mathworks.com/matlabcentral/fileexchange/10526-confidence-region-radius/content/crr.m), March 2008.
5. M. Abramowitz and I. A. Stegun. Handbook of mathematical functions with formulas, graphs, and mathematical tables. Tenth Printing, December 1972, with corrections

15